\documentclass[namedreferences]{solarphysics}
\usepackage[hyperref,optionalrh,solaromanenum]{spr-sola-addons} 
\usepackage{graphicx}                    
\usepackage{natbib}
\usepackage{color}                       
\usepackage{breakurl}                         

\begin{document}

\begin{article}

\begin{opening}

\title{The Solar Radius at 37 GHz through Cycles 22 to 24}

%
\author[addressref={aff1},corref,email={caiuslucius@gmail.com}]{\inits{C.L.}\fnm{Caius L.}~\lnm{Selhorst}\orcid{0000-0002-5897-5236}}
\author[addressref={aff2}]{\inits{J.}\fnm{Juha}~\lnm{Kallunki}\orcid{0000-0002-0325-5565}}
\author[addressref={aff3,aff4}]{\inits{C.G.}\fnm{C. G.}~\lnm{Gim\'enez de Castro}\orcid{0000-0002- 8979-3582}}
\author[addressref={aff3}]{\inits{A.}\fnm{Adriana}~\lnm{Valio}\orcid{0000-0002-1671-8370}}
\author[addressref={aff5}]{\inits{J.E.R.}\fnm{Joaquim E. R.}~\lnm{Costa}\orcid{0000-0002-0703-4735}}

\runningauthor{C. L. Selhorst \it et al.}
\runningtitle{The Solar Radius at 37~GHz}


\address[id=aff1]{NAT - N\'ucleo de Astrof\'isica, Universidade Cruzeiro do Sul/Universidade Cidade de S\~ao Paulo, S\~ao Paulo, SP, Brazil }
\address[id=aff2]{Mets\"ahovi Radio Observatory, Aalto University, Kylm\"al\"a, Finland}
\address[id=aff3]{CRAAM, Universidade Presbiteriana Mackenzie, S\~ao Paulo, SP 01302-907, Brazil}
\address[id=aff4]{IAFE, Universidad de Buenos Aires/CONICET, Buenos Aires, Argentina}
\address[id=aff5]{CEA, Instituto Nacional de Pesquisas Espaciais, S\~ao Jos\'e dos Campos, SP 12201-970, Brazil}

\begin{abstract}
To better understand the influence of the activity cycle on the solar atmosphere, 
we report the time variation of the radius observed at 37~GHz 
($\lambda=8.1$~mm) obtained by the Mets\"ahovi Radio Observatory (MRO) through Solar Cycles 22 to 24 (1989--2015). Almost 5800 maps were analyzed, however, due to instrumental setups changes the data set showed four distinct behaviors, which requested a normalisation process to allow the whole interval analysis. When the whole period was considered, the results showed a positive correlation index of 0.17 between the monthly means of the solar radius at 37~GHz and solar flux obtained at 10.7~cm (F10.7). This correlation index increased to 0.44, when only the data obtained during the last period without instrumental changes were considered (1999--2015). The solar radius correlation with the solar cycle agrees with the previous results obtained at mm/cm wavelengths (17 and 48~GHz), nevertheless, this result is the opposite of that reported at submillimetre wavelengths (212 and 405~GHz).     

\end{abstract}


\end{opening}

 \section{Introduction}
In the last decades, the study of the solar radius variations at different wavelengths and their relation with 
the 11-year cycle has been investigated in many works. 
However, the conclusions for different frequency ranges of the electromagnetic spectrum are still controversial, since 
they depend on the observational method and the wavelength. 

Measurements at photospheric heights using ground-based instruments are inconclusive. Different 
works report a variation of the solar radius correlated with the solar cycle 
\citep{Basu1998,Rozelot1998,Andrei2004,Noel2004,Chapman2008}, whereas other works find the 
opposite behavior and report an anti-correlation of the radius with solar indices
\citep{Gilliland1981,Laclare1996,Kilic2011}. Moreover, \cite{Emilio2005} and \cite{Lefebvre2006} 
report that the optical radius variations have no significant correlation with the solar cycle.

The long useful lifespan of the {\it Solar and Heliospheric Observatory} (SoHO) satellite provides the opportunity to study the temporal 
evolution of the solar radius without the influence of the Earth's atmosphere. \cite{Bush2010} 
analyzed  the optical maps obtained by the {\it Michelson Doppler Imager} (MDI) and reported that 
any intrinsic changes in the solar radius correlated with sunspot cycle must be smaller than 
23~mas  (milli-arcsec) peak to peak. Although the difference is very small, the result is 
larger than the values obtained by the authors in previous works, that determined a variation 
smaller than 15~mas during a whole solar cycle \citep{Emilio2000,Kuhn2004}. 

Moreover, \cite{Kosovichev2018a,Kosovichev2018b} joined 21 years of observations made by MDI/SoHO and HMI/SDO to study the seismic radius at the $f$-mode frequencies. The authors reported a temporal variability of the seismic radius in anti-phase with the solar activity. Since $f$-mode frequencies are not sensitive to temperature or sound-speed variations, the physical mechanism of the inferred displacements is probably associated with magnetic fields accumulating in the subsurface layers during the solar maxima.

At centimeter radio frequencies, variations of the solar radius in phase with the solar cycle 
were shown by \cite{Bachurin1983} at 8~GHz ($\lambda=3.75$~cm) and 13~GHz ($\lambda=2.3$~cm).
\cite{Costa1999} found a similar result at millimeter wavelengths ($\nu=48$~GHz, $\lambda=6.25$~mm). 
\cite{Selhorst2004} and \cite{Selhorst2011} arrived at similar conclusions using NoRH maps at 17~GHz; however, the 
authors point out that the polar radius varies in antiphase. In a recent work, \cite{Menezes2017} analysed a total of {16 600} 
solar maps between 1999 and 2017 and reported a 
strong anticorrelation between the  solar activity and radius 
variation measured at the subterahertz frequencies of 0.212 and 0.405 THz, respectively $1.42$ and $0.74$~mm. For the upper 
atmosphere, \cite{Guigue2007} studied Extreme Ultraviolet Imager maps at $\lambda=304$ 
and $\lambda=171$~{\AA} and found no significant correlation with the solar cycle. 

In this work, we present the temporal evolution of the solar radius at  37~GHz 
($\lambda=8.1$~mm) obtained by the Mets\"ahovi Radio Observatory (MRO) through Solar Cycles 22 to 24 (1989--2015).

\section{Mets\"ahovi Observations} 

\begin{figure}[!htbp]
\centerline{ {\includegraphics[width=12cm]{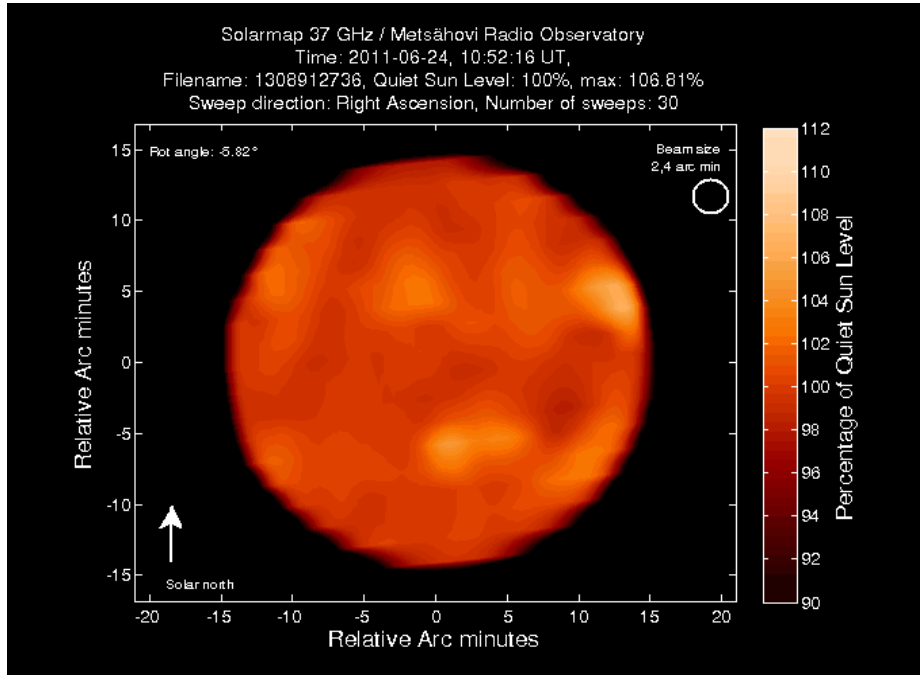}}}
\caption{Solar map obtained by the MRO at 37 GHz on June 24, 2011.}
\label{fig:map}
\end{figure}

The Mets\"ahovi RT-14 telescope at the Mets\"ahovi Radio Observatory (MRO), Aalto University
(Helsinki Region, Finland, GPS: N 60 13.04 E 24 23.35) is a Cassegrain type antenna with
a diameter of 13.7 meters (operated since 1978). The working frequency range of the telescope is 2--150
GHz (13.0 cm -- 2.0 mm). The antenna provides full disk solar mapping, partial solar mapping and,
additionally, the ability to track any selected point on the solar disk. The beam size of the telescope
is $2'.4~(114'')$ at 37 GHz. The receiver is a Dicke type
radiometer, thus the radiometer's own noise will be filtered out. For the temperature stabilisation
of the receiver, a Peltier element is used. The noise temperature of the 37 GHz receiver is around
280 K, and the temporal resolution during the observations is 0.1 s or less. The obtained data are
recorded as intensity and the Quiet Sun temperature is around $8100\pm 300$~K after \cite{Kallunki2018}.

There are two map types in the data set.
Angular size of $50'$ by $33'.6$ with 29 sweeps and 91 samples per sweep in right ascension (RA) and declination (DEC) directions
respectively and $48'$ by $48'$ with 30 sweeps and 200 samples per sweep in RA and DEC directions
respectively. This gives distances between data points of $0.55'~(33'')$ in RA and $1'.16~(69''.5)$ in
DEC direction for the 29 sweep maps. Corresponding distances in the 30 sweep maps are $1'.6~(96'')$
in RA and $0'.24~(14''.4)$ in DEC direction. In the study, only maps scanned in right ascension
were used. Figure~\ref{fig:map} shows an example of a 37 GHz map.

The Mets\"ahovi is operated since 1978, however, only the data digitally recorded were analyzed here, 
that reduce our sample to the data obtained between 1989 and 2015, with almost 5800 maps. 
Along of this period, the instrument setup had tree updates that clearly influenced the solar maps: i) a new radome in 1992; ii) a new main mirror in 1994; iii) a receiver upgrade in 1997.  These instrumental changes clearly affected the results, nevertheless, the data time serie, are still suitable for the proposed study. 
  
\section{Data Analysis and Results}

\begin{figure}[!htbp]
\centerline{ {\includegraphics[width=12cm]{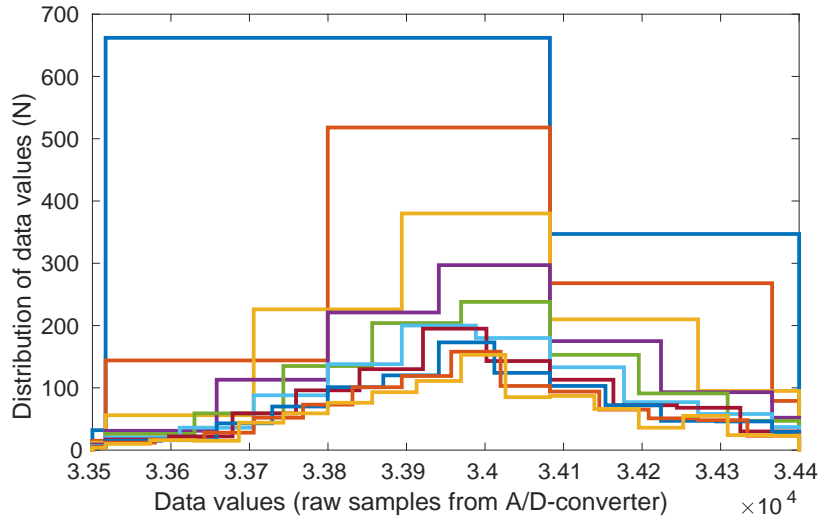}}}
\caption{Histogram of the raw solar intensity measurements with different bin sizes.}
\label{fig:histo}
\end{figure}

To determine the solar disk radius, a process similar to the one defined in \cite{Costa1999} was applied.
The Quiet Sun Level (QSL) was determined by taking the average reading of the maximum values
of histograms made with 30 different bin sizes. This yield the QSL as the most common brightness
temperature on a map. Figure~\ref{fig:histo} shows an example of the 30 different bin size 
histograms used to determine the QSL. 

\begin{figure}[!htbp] 
\centerline{ {\includegraphics[width=12cm]{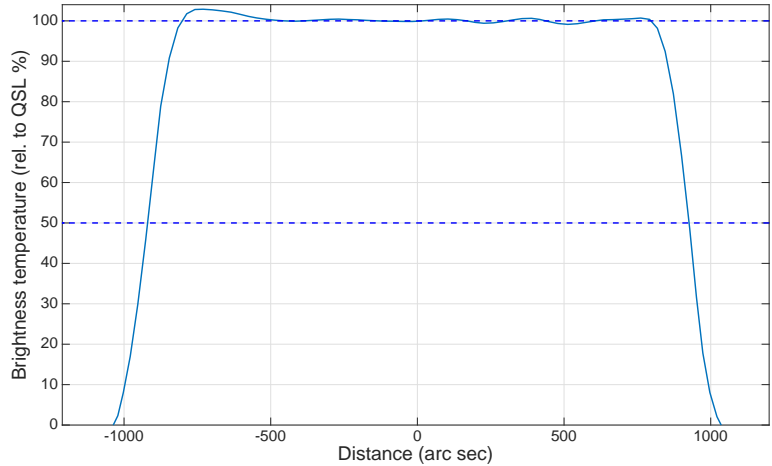}}}
\caption{Example of a normalised intensity profile of a map sweep. The dashed lines are the QSL and 50~\% QSL levels.}
\label{fig:QSL}
\end{figure}

The solar limb was assumed at 50~\% of the QSL. The coordinates of the solar limb
were determined in two directions, right ascension and declination. The limb coordinates were
linearly interpolated from the values closest to 50~\% QSL. An example 
of a normalised intensity profile is shown in Figure~\ref{fig:QSL} , in which the dashed lines are the QSL and 50~\% QSL levels. Limb 
positions were available on
varying intervals depending on whether the direction of limb detection was right ascension or
declination. Also, Mets\"ahovi's dataset includes solar maps with varying number of sweeps and
samples per sweep. 

A least squares circle fit was made using the acquired limb coordinates. Limb positions
deviating more than $0'.3~(18'')$ from the fitted radius were discarded. Then, a new circle was adjusted to the remaining points, and again the limb points with deviations greater than 0'.3 were discarded. This process was recursively repeated  until the difference of subsequent radius values was less than $0'.0005 (0''.03)$ or until 20 iterations had been reached.

\begin{figure}[!htbp]
\centerline{ {\includegraphics[width=12cm]{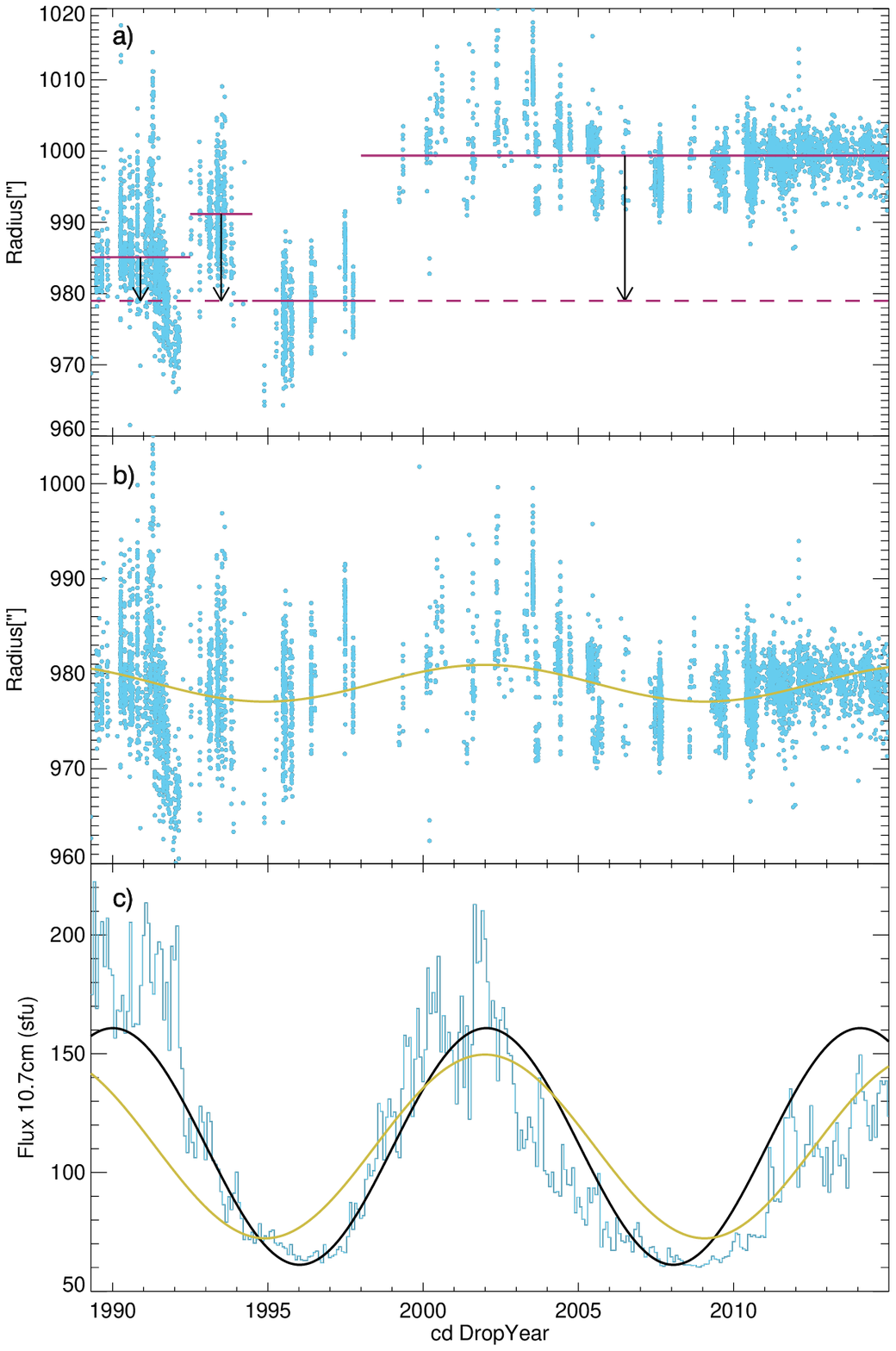}}}
\caption{a) Solar radius obtained with distinct MRO instrumental configurations. The continuous horizontal magenta lines are the mean values obtained for each period, while the arrows represent the subtracted values. b) The mean radii normalised by the smaller mean radius listed in Table~\ref{table1}.  c) The averaged monthly solar flux obtained at 10.7~cm. The yellow curves on b) and c) have the period adjusted to the MRO data, whereas the black curve is a fit to the solar flux at 10.7~cm.}
\label{fig:radius}
\end{figure}

The results clearly show the influence of the instrumental changes (Figure~\ref{fig:radius}a). The averaged radii of each period without instrumental changes are listed in Table~\ref{table1}. In which the first dataset was obtained with the original instrumentation,  the second one was obtained after the radome replacement in 1992, the third dataset was acquired after the main mirror substitution in 1994, and last dataset was obtained after installation of the new receiver in 1997. Since the solar radii increased after the radome change but decreased after the main mirror substitution, the observed radii increment may not have been caused by the radome itself, but due to  some degradation of the old main mirror. 

Following \cite{Battaglia2017}, two sets of errors were considered with respect to the averaged solar radius. The first one was associated to the circular fit of the limb points, that can be affected by several issues like the beam pattern, the weather condition, and the yearly variation  of the Sun's elevation angle during the observation. The second error was caused by the solar atmospheric variation along the cycles. The smaller mean radius  $979''\pm 5'' \pm 6''$ was obtained during the period after the update of the main mirror. Nevertheless, the receiver upgrade resulted in an increase of $20''$ in the averaged value, namely $999''\pm4''\pm 5''$. This radius increase probably reflects a larger beam pattern of the new receiver, while the smaller errors were caused by the low activity period and the greater number of measurements obtained in the last period. 

Since the main purpose of this work is the analysis of the variation of the 37~GHz radius along the solar cycle, the solar radii values obtained at distinct periods with different instrumental setups were normalised to the minimum mean value obtained ($979''$), {\it i.e.}, the differences between averaged values were subtracted (Figure~\ref{fig:radius}b). This value agrees with the solar radii observations at mm/cm compiled by \cite{Rozelot2015}.    

\begin{table}
\caption{Solar radius.}
\label{table1}
\begin{tabular}{ccc}
\hline
\hline
Observed period & Number of maps &Averaged radius \\
\hline
April 1989 -- April 1992 & 1104 & $985''\pm5''\pm 8''$ \\
July 1992 -- April 1994& 392 & $991''\pm 7'' \pm 7''$ \\
November 1994 -- October 1997 & 446 & $979''\pm 5'' \pm 6''$ \\
March 1999 -- December 2014 & 3847 & $999''\pm 4'' \pm 5''$\\
\hline
\end{tabular} 
\end{table}  

After the data normalisation (Figure~\ref{fig:radius}b), a Lomb-Scargle periodogram \citep{Scargle1982} was used to  investigate the influence of the solar cycle on the radii variation yielding a period of 14.2 years with a significance index greater than 99.99\% (yellow curve in Figure~\ref{fig:radius}b).  The central maxima of the yellow curve in Figure~\ref{fig:radius}b was placed on the ending of 2001. 

To compare the radius with the solar cycle we used   the averaged monthly solar flux at 10.7~cm (F10.7),  plotted on Figure~\ref{fig:radius}c. The Lomb-Scargle analysis of F10.7  showed a 12.0 years period  with a significance index greater than 99.99\% (black curve in Figure~\ref{fig:radius}c), with the central maxima coinciding with the one obtained in the radius analysis.    

The black curve in Figure~\ref{fig:radius}c has a better agreement with the data obtained before 2002, which could reflect the unusual prolonged solar minimum observed after the central maxima. On the other hand, the longer period period obtained in the radii analyses could be caused by the changes in the instrumental settings that occurred in the studied period.    

\begin{figure}[!htbp]
\centerline{ {\includegraphics[width=12cm]{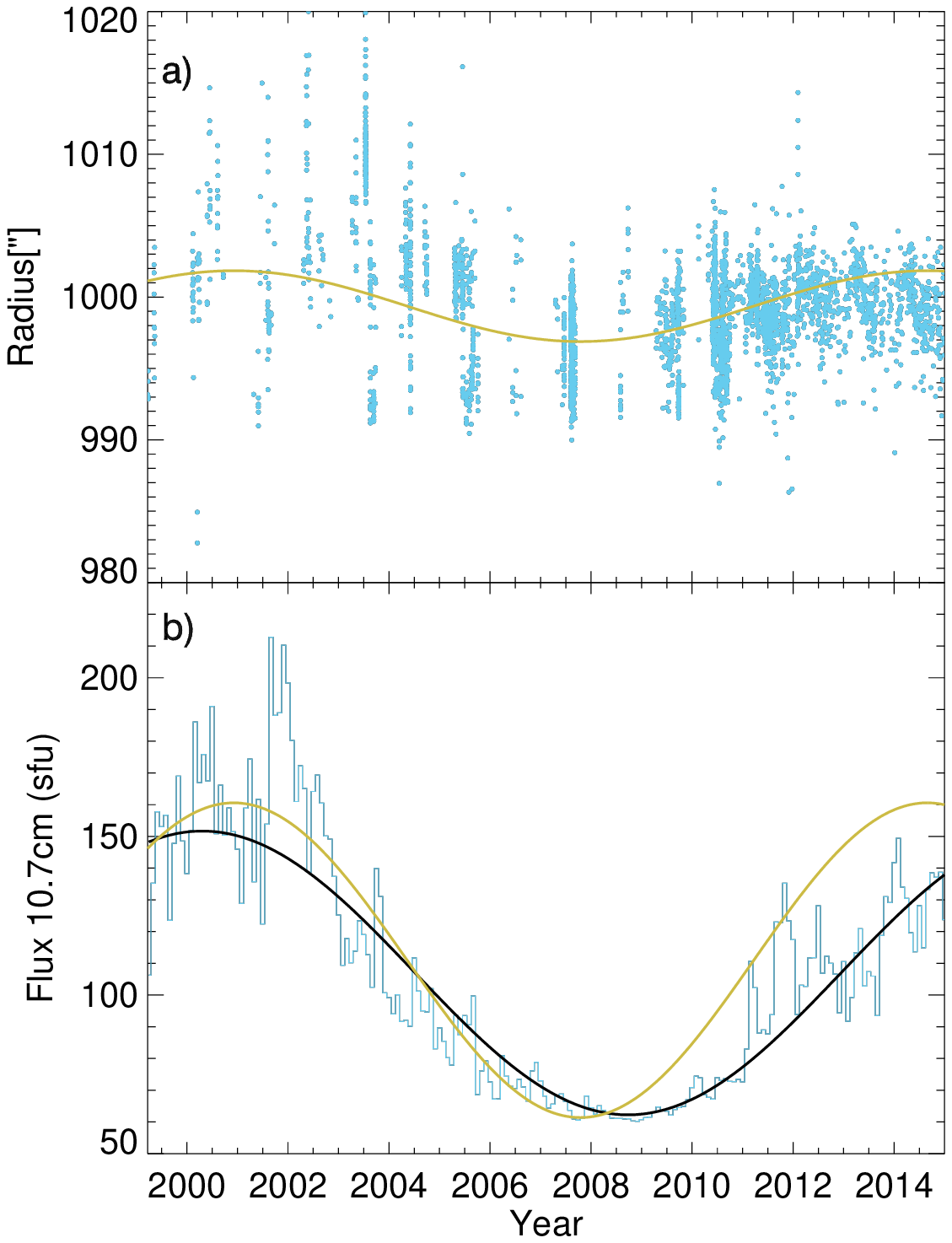}}}
\caption{a) Solar radius obtained after March 1999, that kept the same MRO instrumental configuration.   b) The averaged monthly solar flux obtained at 10.7~cm. The yellow curves on b) and c) have the period adjusted to the MRO data, whereas the black curve was fit to the solar flux at 10.7~cm.}
\label{fig:radius2}
\end{figure}

Trying to avoid the effect of the instrumental changes, we re-analyzed the radius variation considering only the last dataset, {\it i.e.}, the radii obtained after March 1999 (Figure~\ref{fig:radius2}a). The results show a small reduction in the previous obtained period, that changes from 14.2 to 13.7 years, nevertheless, the F10.7 showed an increase in this period, that varies from 12.0 to 16.9 years. As can be seen in Figure~\ref{fig:radius2}b, the F10.7 fit reflects the unusual prolonged solar minimum.   

To verify the degree of correlation between the samples, we used the Pearson linear test. For the whole interval (1989--2015), the correlation index between the monthly F10.7 and the averaged monthly radii was 0.17. Considering the 176 months with radius measurements, the probability of the correlation between the radii and F10.7  is 97.7\%.  On the other hand, for the data obtained with the last instrumental settings, {\it i.e.}, after March 1999, the number of months with measurements is reduced to 117 and the correlation with the solar cycle increases to 0.44, yielding a 100\% correlation probability between the datasets.  

\section{Discussion and Conclusions}

In this work, we studied the solar radius variation using the solar maps obtained by the Mets\"ahovi RT-14 telescope at the Mets\"ahovi Radio Observatory (MRO). The analyzed data consist of almost 5800 maps digitally recorded, that were obtained between 1989 and 2015. Along this period, the instrument received three important updates: i) a new radome in 1992; ii) a new main mirror in 1994; and iii) a receiver upgrade in 1997. While the smallest mean solar radius was obtained after the installation of the new main mirror, the replacement of the receiver caused an increase of approximately  $20''$ in the average radius.

Radio observation are affected by several distortions (beam pattern, weather conditions, Sun's elevation changes along 
the year, and other parameters).  These requirements are reflected in the variation of the limb points around the circle 
fit, and can be considered as systematic errors. Their effects are difficult to quantify, and they do not go away by merely adding observations to enfeeble their proper impact.
The way we proceed facing the instrumental changes along the years have been regular and uniform, which enable us to infer that
results obtained here are consistent with the previous studies at longer wavelengths. A more deeper error analysis could be conducted in further works.

As shown in Figure~\ref{fig:radius}a, the instrumental changes clearly influenced the radii values, which can be due to changes in the antenna beamwidth or in the beam sidelobes. However, we do not have enough information to recover these beam profiles and evaluate their influence. Each distinct period of same instrumental configuration yield different mean radii values (see Table~\ref{table1}), from $979''\pm 5''\pm 6''$ to $999''\pm4''\pm 5''$, determined in the last and longest interval without instrumental changes. The minimum mean value agrees with the solar radii observations at mm/cm compiled by \cite{Rozelot2015}, whereas, the other mean values are greater than that expected at 37~GHz. 

Since the main purpose of this work is the analysis of the variation of the 37 GHz solar radius along the solar cycle, we merged the radii obtained under distinct instrumental setups and normalised these values to the minimum calculated mean radius (Figure~\ref{fig:radius}b). After the radii normalisation, a Lomb-Scargle periodogram \citep{Scargle1982}, resulted in a 14.2 years periodic variation, which is greater than the 12.0 years period obtained from the F10.7. The discrepancy between the obtained periods can be addressed to the unusual long minimum between cycles 23 and 24. Moreover, the correlation index between the monthly F10.7 and the averaged monthly radii was 0.17. Nevertheless, when considering only the data obtained after March 1999, {\it i.e.}, the  correlation between the radius variation and the  F10.7 increased to 0.44. 

The correlation between the solar cycle and solar radius variation obtained at 37~GHz agrees with the previous results obtained at 17 and 48~GHz \citep{Selhorst2004,Selhorst2011,Costa1999}. However, there is a discrepancy with results reported by \cite{Menezes2017} at submillimetre wavelengths (212 and 405~GHz),  that agrees with seismic radius variation observed at the $f$-mode reported by \cite{Kosovichev2018a,Kosovichev2018b}. These apparent contradictions at the radio wavelengths may be explained by photospheric contribution to the submillimetric emission (Figure 4 of \cite{Selhorst2019}), that could reflect the seismic radius behavior. Nevertheless, the seismic variation seems to vanish at the smaller radio frequencies that are formed at higher chromospheric altitudes. 

\begin{acks}
C.L.S. acknowledges financial support from the S\~ao Paulo Research Foundation (FAPESP), grant number 2019/ 03301-8. C. L. S. and C. G. G. C. are thankful to CNPq by the support through grants 306638/2018-5 and 305203/2016-9, respectively. The authors also thank to Niko Lavonen for help with the data reduction.
\end{acks}

\begin{acks}[Disclosure of Potential Conflicts of Interest]  The authors declare that they have no conflicts of interest. \end{acks}

\bibliographystyle{spr-mp-sola}
\bibliography{Selhorst_rev2}  

\begin{thebibliography}{26}
\ifx\bisbn     \undefined \def\bisbn  #1{ISBN #1}\fi
\ifx\binits    \undefined \def\binits#1{#1}\fi
\ifx\bauthor   \undefined \def\bauthor#1{#1}\fi
\ifx\batitle   \undefined \def\batitle#1{#1}\fi
\ifx\bjtitle   \undefined \def\bjtitle#1{\textit{#1}}\fi
\ifx\bvolume   \undefined \def\bvolume#1{\textbf{#1}}\fi
\ifx\byear     \undefined \def\byear#1{#1}\fi
\ifx\bissue    \undefined \def\bissue#1{#1}\fi
\ifx\bfpage    \undefined \def\bfpage#1{#1}\fi
\ifx\blpage    \undefined \def\blpage #1{#1}\fi
\ifx\burl      \undefined \def\burl#1{\textsf{#1}}\fi
\ifx\href      \undefined \def\href#1#2{\textsf{#2}}\fi
\ifx\betal     \undefined \def\betal{\textit{et al.}}\fi
\ifx\bctitle   \undefined \def\bctitle#1{#1}\fi
\ifx\beditor   \undefined \def\beditor#1{#1}\fi
\ifx\bbtitle   \undefined \def\bbtitle#1{\textit{#1}}\fi
\ifx\bedition  \undefined \def\bedition#1{#1}\fi
\ifx\bseriesno \undefined \def\bseriesno#1{\textbf{#1}}\fi
\ifx\blocation \undefined \def\blocation#1{#1}\fi
\ifx\bsertitle \undefined \def\bsertitle#1{\textit{#1}}\fi
\ifx\bsnm      \undefined \def\bsnm#1{#1}\fi
\ifx\bsuffix   \undefined \def\bsuffix#1{#1}\fi
\ifx\bparticle \undefined \def\bparticle#1{#1}\fi
\ifx\barticle  \undefined \def\barticle#1{}\fi
\ifx\binstitute  \undefined \def\binstitute#1{#1}\fi
\ifx\bpublisher  \undefined \def\bpublisher#1{#1}\fi
\ifx\doiurl    \undefined \def\doiurl#1{\href{#1}{\textsf{DOI}}}\fi
\makeatletter
\def\safeHref#1#2#3{\in@{http}{#2}\ifin@\href{#2}{#3}\else\href{#1#2}{#3}\fi}
\makeatother
\ifx\adsurl    \undefined
  \def\adsurl#1{\safeHref{http://adsabs.harvard.edu/abs/}{#1}{\textsf{ADS}}}\fi
\ifx\arxivurl  \undefined
  \def\arxivurl#1{\safeHref{http://arxiv.org/abs/}{#1}{\textsf{arXiv}}}\fi
\ifx\botherref \undefined \def\botherref#1{}\fi
\ifx\url       \undefined \def\url#1{\textsf{#1}}\fi
\ifx\bchapter  \undefined \def\bchapter#1{}\fi
\ifx\bbook     \undefined \def\bbook#1{}\fi
\ifx\bcomment  \undefined \def\bcomment#1{#1}\fi
\ifx\oauthor   \undefined \def\oauthor#1{#1}\fi
\ifx\citeauthoryear \undefined\def \citeauthoryear#1{#1}\fi
\def\endbibitem {}
\ifx\bconflocation  \undefined \def\bconflocation#1{#1} \fi

\bibitem[\protect\citeauthoryear{{Andrei} \textit{et~al.}}{2004}]{Andrei2004}
\begin{barticle}
\bauthor{\bsnm{{Andrei}}, \binits{A.H.}},
\bauthor{\bsnm{{Boscardin}}, \binits{S.C.}},
\bauthor{\bsnm{{Chollet}}, \binits{F.}},
\bauthor{\bsnm{{Delmas}}, \binits{C.}},
\bauthor{\bsnm{{Golbasi}}, \binits{O.}},
\bauthor{\bsnm{{Jilinski}}, \binits{E.G.}},
\bauthor{\bsnm{{Kili{\c c}}}, \binits{H.}},
\bauthor{\bsnm{{Laclare}}, \binits{F.}},
\bauthor{\bsnm{{Morand}}, \binits{F.}},
\bauthor{\bsnm{{Penna}}, \binits{J.L.}},
\bauthor{\bsnm{{Reis Neto}}, \binits{E.}}:
\byear{2004},
\batitle{{Comparison of CCD astrolabe multi-site solar diameter observations}}.
\bjtitle{\aap}
\bvolume{427},
\bfpage{717}.
\doiurl{https://doi.org/10.1051/0004-6361:20041334}.
\adsurl{2004A\%26A...427..717A}.
\end{barticle}
\endbibitem

\bibitem[\protect\citeauthoryear{{Bachurin}}{1983}]{Bachurin1983}
\begin{barticle}
\bauthor{\bsnm{{Bachurin}}, \binits{A.F.}}:
\byear{1983},
\batitle{{Variation in solar radio radius with the phase of the solar activity
  cycle at wavelengths of 2.25 and 3.5 CM}}.
\bjtitle{Izvestiya Ordena Trudovogo Krasnogo Znameni Krymskoj Astrofizicheskoj
  Observatorii}
\bvolume{68},
\bfpage{68}.
\adsurl{1983IzKry..68...68B}.
\end{barticle}
\endbibitem

\bibitem[\protect\citeauthoryear{{Basu}}{1998}]{Basu1998}
\begin{barticle}
\bauthor{\bsnm{{Basu}}, \binits{D.}}:
\byear{1998},
\batitle{{Radius of the Sun in relation to solar activity}}.
\bjtitle{\solphys}
\bvolume{183},
\bfpage{291}.
\adsurl{1998SoPh..183..291B}.
\end{barticle}
\endbibitem

\bibitem[\protect\citeauthoryear{{Battaglia}
  \textit{et~al.}}{2017}]{Battaglia2017}
\begin{barticle}
\bauthor{\bsnm{{Battaglia}}, \binits{M.}},
\bauthor{\bsnm{{Hudson}}, \binits{H.S.}},
\bauthor{\bsnm{{Hurford}}, \binits{G.J.}},
\bauthor{\bsnm{{Krucker}}, \binits{S.}},
\bauthor{\bsnm{{Schwartz}}, \binits{R.A.}}:
\byear{2017},
\batitle{{The solar x-ray limb}}.
\bjtitle{\apj}
\bvolume{843}(\bissue{2}),
\bfpage{123}.
\doiurl{https://doi.org/10.3847/1538-4357/aa76da}.
\adsurl{https://ui.adsabs.harvard.edu/abs/2017ApJ...843..123B}.
\end{barticle}
\endbibitem

\bibitem[\protect\citeauthoryear{{Bush}, {Emilio}, and {Kuhn}}{2010}]{Bush2010}
\begin{barticle}
\bauthor{\bsnm{{Bush}}, \binits{R.I.}},
\bauthor{\bsnm{{Emilio}}, \binits{M.}},
\bauthor{\bsnm{{Kuhn}}, \binits{J.R.}}:
\byear{2010},
\batitle{{On the constancy of the solar radius. III.}}
\bjtitle{\apj}
\bvolume{716},
\bfpage{1381}.
\doiurl{https://doi.org/10.1088/0004-637X/716/2/1381}.
\adsurl{2010ApJ...716.1381B}.
\end{barticle}
\endbibitem

\bibitem[\protect\citeauthoryear{{Chapman}, {Dobias}, and
  {Walton}}{2008}]{Chapman2008}
\begin{barticle}
\bauthor{\bsnm{{Chapman}}, \binits{G.A.}},
\bauthor{\bsnm{{Dobias}}, \binits{J.J.}},
\bauthor{\bsnm{{Walton}}, \binits{S.R.}}:
\byear{2008},
\batitle{{On the variability of the apparent solar radius}}.
\bjtitle{\apj}
\bvolume{681},
\bfpage{1698}.
\doiurl{https://doi.org/10.1086/588512}.
\adsurl{2008ApJ...681.1698C}.
\end{barticle}
\endbibitem

\bibitem[\protect\citeauthoryear{{Costa} \textit{et~al.}}{1999}]{Costa1999}
\begin{barticle}
\bauthor{\bsnm{{Costa}}, \binits{J.E.R.}},
\bauthor{\bsnm{{Silva}}, \binits{A.V.R.}},
\bauthor{\bsnm{{Makhmutov}}, \binits{V.S.}},
\bauthor{\bsnm{{Rolli}}, \binits{E.}},
\bauthor{\bsnm{{Kaufmann}}, \binits{P.}},
\bauthor{\bsnm{{Magun}}, \binits{A.}}:
\byear{1999},
\batitle{{Solar radius variations at 48 GHz correlated with solar irradiance}}.
\bjtitle{\apjl}
\bvolume{520},
\bfpage{L63}.
\doiurl{https://doi.org/10.1086/312132}.
\adsurl{http://ads.astro.puc.cl/abs/1999ApJ...520L..63C}.
\end{barticle}
\endbibitem

\bibitem[\protect\citeauthoryear{{Emilio} and {Leister}}{2005}]{Emilio2005}
\begin{barticle}
\bauthor{\bsnm{{Emilio}}, \binits{M.}},
\bauthor{\bsnm{{Leister}}, \binits{N.V.}}:
\byear{2005},
\batitle{{Solar diameter measurements at S{\~a}o Paulo Observatory}}.
\bjtitle{\mnras}
\bvolume{361},
\bfpage{1005}.
\doiurl{https://doi.org/10.1111/j.1365-2966.2005.09236.x}.
\adsurl{2005MNRAS.361.1005E}.
\end{barticle}
\endbibitem

\bibitem[\protect\citeauthoryear{{Emilio} \textit{et~al.}}{2000}]{Emilio2000}
\begin{barticle}
\bauthor{\bsnm{{Emilio}}, \binits{M.}},
\bauthor{\bsnm{{Kuhn}}, \binits{J.R.}},
\bauthor{\bsnm{{Bush}}, \binits{R.I.}},
\bauthor{\bsnm{{Scherrer}}, \binits{P.}}:
\byear{2000},
\batitle{{On the constancy of the solar diameter}}.
\bjtitle{\apj}
\bvolume{543},
\bfpage{1007}.
\doiurl{https://doi.org/10.1086/317157}.
\adsurl{2000ApJ...543.1007E}.
\end{barticle}
\endbibitem

\bibitem[\protect\citeauthoryear{{Gilliland}}{1981}]{Gilliland1981}
\begin{barticle}
\bauthor{\bsnm{{Gilliland}}, \binits{R.L.}}:
\byear{1981},
\batitle{{Solar radius variations over the past 265 years}}.
\bjtitle{\apj}
\bvolume{248},
\bfpage{1144}.
\doiurl{https://doi.org/10.1086/159243}.
\adsurl{1981ApJ...248.1144G}.
\end{barticle}
\endbibitem

\bibitem[\protect\citeauthoryear{{Gim{\'e}nez de Castro}
  \textit{et~al.}}{2007}]{Guigue2007}
\begin{barticle}
\bauthor{\bsnm{{Gim{\'e}nez de Castro}}, \binits{C.G.}},
\bauthor{\bsnm{{Varela Saraiva}}, \binits{A.C.}},
\bauthor{\bsnm{{Costa}}, \binits{J.E.R.}},
\bauthor{\bsnm{{Selhorst}}, \binits{C.L.}}:
\byear{2007},
\batitle{{The solar radius in the EUV during the cycle XXIII}}.
\bjtitle{\aap}
\bvolume{476},
\bfpage{369}.
\doiurl{https://doi.org/10.1051/0004-6361:20078118}.
\adsurl{2007A\%26A...476..369G}.
\end{barticle}
\endbibitem

\bibitem[\protect\citeauthoryear{{Kallunki} and
  {Tornikoski}}{2018}]{Kallunki2018}
\begin{barticle}
\bauthor{\bsnm{{Kallunki}}, \binits{J.}},
\bauthor{\bsnm{{Tornikoski}}, \binits{M.}}:
\byear{2018},
\batitle{{Measurements of the quiet-Sun level brightness temperature at 8 mm}}.
\bjtitle{\solphys}
\bvolume{293},
\bfpage{156}.
\doiurl{https://doi.org/10.1007/s11207-018-1380-8}.
\adsurl{2018SoPh..293..156K}.
\end{barticle}
\endbibitem

\bibitem[\protect\citeauthoryear{{Kilic} and {Golbasi}}{2011}]{Kilic2011}
\begin{barticle}
\bauthor{\bsnm{{Kilic}}, \binits{H.}},
\bauthor{\bsnm{{Golbasi}}, \binits{O.}}:
\byear{2011},
\batitle{{Comparison of long-term trend of solar radius with sunspot activity
  and flare index}}.
\bjtitle{\apss}
\bvolume{334},
\bfpage{75}.
\doiurl{https://doi.org/10.1007/s10509-011-0714-x}.
\adsurl{2011Ap\%26SS.334...75K}.
\end{barticle}
\endbibitem

\bibitem[\protect\citeauthoryear{{Kosovichev} and
  {Rozelot}}{2018a}]{Kosovichev2018a}
\begin{barticle}
\bauthor{\bsnm{{Kosovichev}}, \binits{A.}},
\bauthor{\bsnm{{Rozelot}}, \binits{J.-P.}}:
\byear{2018}a,
\batitle{{Cyclic changes of the Sun{\textquoteright}s seismic radius}}.
\bjtitle{\apj}
\bvolume{861}(\bissue{2}),
\bfpage{90}.
\doiurl{https://doi.org/10.3847/1538-4357/aac81d}.
\adsurl{https://ui.adsabs.harvard.edu/abs/2018ApJ...861...90K}.
\end{barticle}
\endbibitem

\bibitem[\protect\citeauthoryear{{Kosovichev} and
  {Rozelot}}{2018b}]{Kosovichev2018b}
\begin{barticle}
\bauthor{\bsnm{{Kosovichev}}, \binits{A.G.}},
\bauthor{\bsnm{{Rozelot}}, \binits{J.P.}}:
\byear{2018}b,
\batitle{{Solar cycle variations of rotation and asphericity in the
  near-surface shear layer}}.
\bjtitle{Journal of Atmospheric and Solar-Terrestrial Physics}
\bvolume{176},
\bfpage{21}.
\doiurl{https://doi.org/10.1016/j.jastp.2017.08.004}.
\adsurl{https://ui.adsabs.harvard.edu/abs/2018JASTP.176...21K}.
\end{barticle}
\endbibitem

\bibitem[\protect\citeauthoryear{{Kuhn} \textit{et~al.}}{2004}]{Kuhn2004}
\begin{barticle}
\bauthor{\bsnm{{Kuhn}}, \binits{J.R.}},
\bauthor{\bsnm{{Bush}}, \binits{R.I.}},
\bauthor{\bsnm{{Emilio}}, \binits{M.}},
\bauthor{\bsnm{{Scherrer}}, \binits{P.H.}}:
\byear{2004},
\batitle{{On the constancy of the solar diameter. II.}}
\bjtitle{\apj}
\bvolume{613},
\bfpage{1241}.
\doiurl{https://doi.org/10.1086/423301}.
\adsurl{2004ApJ...613.1241K}.
\end{barticle}
\endbibitem

\bibitem[\protect\citeauthoryear{{Laclare} \textit{et~al.}}{1996}]{Laclare1996}
\begin{barticle}
\bauthor{\bsnm{{Laclare}}, \binits{F.}},
\bauthor{\bsnm{{Delmas}}, \binits{C.}},
\bauthor{\bsnm{{Coin}}, \binits{J.P.}},
\bauthor{\bsnm{{Irbah}}, \binits{A.}}:
\byear{1996},
\batitle{{Measurements and variations of the solar diameter}}.
\bjtitle{\solphys}
\bvolume{166},
\bfpage{211}.
\doiurl{https://doi.org/10.1007/BF00149396}.
\adsurl{1996SoPh..166..211L}.
\end{barticle}
\endbibitem

\bibitem[\protect\citeauthoryear{{Lefebvre}
  \textit{et~al.}}{2006}]{Lefebvre2006}
\begin{barticle}
\bauthor{\bsnm{{Lefebvre}}, \binits{S.}},
\bauthor{\bsnm{{Bertello}}, \binits{L.}},
\bauthor{\bsnm{{Ulrich}}, \binits{R.K.}},
\bauthor{\bsnm{{Boyden}}, \binits{J.E.}},
\bauthor{\bsnm{{Rozelot}}, \binits{J.P.}}:
\byear{2006},
\batitle{{Solar radius measurements at Mount Wilson Observatory}}.
\bjtitle{\apj}
\bvolume{649},
\bfpage{444}.
\doiurl{https://doi.org/10.1086/506134}.
\adsurl{2006ApJ...649..444L}.
\end{barticle}
\endbibitem

\bibitem[\protect\citeauthoryear{{Menezes} and {Valio}}{2017}]{Menezes2017}
\begin{barticle}
\bauthor{\bsnm{{Menezes}}, \binits{F.}},
\bauthor{\bsnm{{Valio}}, \binits{A.}}:
\byear{2017},
\batitle{{Solar radius at subterahertz frequencies and its relation to solar
  activity}}.
\bjtitle{\solphys}
\bvolume{292},
\bfpage{id. 195}.
\doiurl{https://doi.org/10.1007/s11207-017-1216-y}.
\adsurl{http://cdsads.u-strasbg.fr/abs/2017SoPh..292..195M}.
\end{barticle}
\endbibitem

\bibitem[\protect\citeauthoryear{{No{\"e}l}}{2004}]{Noel2004}
\begin{barticle}
\bauthor{\bsnm{{No{\"e}l}}, \binits{F.}}:
\byear{2004},
\batitle{{Solar cycle dependence of the apparent radius of the Sun}}.
\bjtitle{\aap}
\bvolume{413},
\bfpage{725}.
\doiurl{https://doi.org/10.1051/0004-6361:20031573}.
\adsurl{2004A\%26A...413..725N}.
\end{barticle}
\endbibitem

\bibitem[\protect\citeauthoryear{{Rozelot}}{1998}]{Rozelot1998}
\begin{barticle}
\bauthor{\bsnm{{Rozelot}}, \binits{J.P.}}:
\byear{1998},
\batitle{{The correlation between the Calern solar diameter measurements and
  the solar irradiance}}.
\bjtitle{\solphys}
\bvolume{177},
\bfpage{321}.
\adsurl{1998SoPh..177..321R}.
\end{barticle}
\endbibitem

\bibitem[\protect\citeauthoryear{{Rozelot}, {Kosovichev}, and
  {Kilcik}}{2015}]{Rozelot2015}
\begin{barticle}
\bauthor{\bsnm{{Rozelot}}, \binits{J.P.}},
\bauthor{\bsnm{{Kosovichev}}, \binits{A.}},
\bauthor{\bsnm{{Kilcik}}, \binits{A.}}:
\byear{2015},
\batitle{{Solar radius variations: an inquisitive wavelength dependence}}.
\bjtitle{\apj}
\bvolume{812},
\bfpage{91}.
\doiurl{https://doi.org/10.1088/0004-637X/812/2/91}.
\adsurl{http://cdsads.u-strasbg.fr/abs/2015ApJ...812...91R}.
\end{barticle}
\endbibitem

\bibitem[\protect\citeauthoryear{{Scargle}}{1982}]{Scargle1982}
\begin{barticle}
\bauthor{\bsnm{{Scargle}}, \binits{J.D.}}:
\byear{1982},
\batitle{{Studies in astronomical time series analysis. II - Statistical
  aspects of spectral analysis of unevenly spaced data}}.
\bjtitle{\apj}
\bvolume{263},
\bfpage{835}.
\doiurl{https://doi.org/10.1086/160554}.
\adsurl{1982ApJ...263..835S}.
\end{barticle}
\endbibitem

\bibitem[\protect\citeauthoryear{{Selhorst}, {Silva}, and
  {Costa}}{2004}]{Selhorst2004}
\begin{barticle}
\bauthor{\bsnm{{Selhorst}}, \binits{C.L.}},
\bauthor{\bsnm{{Silva}}, \binits{A.V.R.}},
\bauthor{\bsnm{{Costa}}, \binits{J.E.R.}}:
\byear{2004},
\batitle{{Radius variations over a solar cycle}}.
\bjtitle{\aap}
\bvolume{420},
\bfpage{1117}.
\adsurl{cgi-bin/nph-bib_query?bibcode=2004A\%26A...420.1117S&amp;db_key=AST}.
\end{barticle}
\endbibitem

\bibitem[\protect\citeauthoryear{{Selhorst}
  \textit{et~al.}}{2011}]{Selhorst2011}
\begin{barticle}
\bauthor{\bsnm{{Selhorst}}, \binits{C.L.}},
\bauthor{\bsnm{{Gim{\'e}nez de Castro}}, \binits{C.G.}},
\bauthor{\bsnm{{V{\'a}lio}}, \binits{A.}},
\bauthor{\bsnm{{Costa}}, \binits{J.E.R.}},
\bauthor{\bsnm{{Shibasaki}}, \binits{K.}}:
\byear{2011},
\batitle{{The behavior of the 17 GHz solar radius and limb brightening in the
  spotless minimum XXIII/XXIV}}.
\bjtitle{\apj}
\bvolume{734},
\bfpage{64}.
\doiurl{https://doi.org/10.1088/0004-637X/734/1/64}.
\adsurl{2011ApJ...734...64S}.
\end{barticle}
\endbibitem

\bibitem[\protect\citeauthoryear{{Selhorst}
  \textit{et~al.}}{2019}]{Selhorst2019}
\begin{barticle}
\bauthor{\bsnm{{Selhorst}}, \binits{C.L.}},
\bauthor{\bsnm{{Sim{\~o}es}}, \binits{P.J.A.}},
\bauthor{\bsnm{{Braj{\v{s}}a}}, \binits{R.}},
\bauthor{\bsnm{{Valio}}, \binits{A.}},
\bauthor{\bsnm{{Gim{\'e}nez de Castro}}, \binits{C.G.}},
\bauthor{\bsnm{{Costa}}, \binits{J.E.R.}},
\bauthor{\bsnm{{Menezes}}, \binits{F.}},
\bauthor{\bsnm{{Rozelot}}, \binits{J.P.}},
\bauthor{\bsnm{{Hales}}, \binits{A.S.}},
\bauthor{\bsnm{{Iwai}}, \binits{K.}},
\bauthor{\bsnm{{White}}, \binits{S.}}:
\byear{2019},
\batitle{{Solar polar brightening and radius at 100 and 230 GHz observed by
  ALMA}}.
\bjtitle{\apj}
\bvolume{871}(\bissue{1}),
\bfpage{45}.
\doiurl{https://doi.org/10.3847/1538-4357/aaf4f2}.
\adsurl{https://ui.adsabs.harvard.edu/abs/2019ApJ...871...45S}.
\end{barticle}
\endbibitem

\end{thebibliography}

\end{article} 
\end{document}